\documentclass[3p, 12pt, onecolumn]{elsarticle}
\usepackage[T1]{fontenc}
\usepackage{amssymb}
\usepackage{caption}
\usepackage{subcaption}
\usepackage{graphicx,wrapfig,lipsum}
\usepackage{amscd,amsmath,verbatim}
\usepackage{amsfonts,epsfig}
\usepackage{mathptmx}
\usepackage{setspace}
\usepackage{epsfig}
\usepackage{url}
\usepackage{color}

\pdfoutput=1

\begin{document}

\journal{Physica A}

\begin{frontmatter}

\title{Exploring Crossing Times and Congestion Patterns at Scramble Intersections in Pedestrian Dynamics Models: A Statistical Analysis}

\author{Eduardo V. Stock, Roberto da Silva} 

\address{Instituto de F\'{i}sica, Universidade Federal do Rio Grande do Sul,
Porto Alegre, Rio Grande do Sul, Brazil\\
}

\begin{abstract}

Scramble intersections stand as compelling examples of complex systems, shedding light 
on the pressing challenge of urban mobility. In this paper, we introduce a model aimed at 
unraveling the statistical intricacies of pedestrian crossing times and their fluctuations in 
scenarios commonly encountered in major urban centers. Our findings offer snapshots that faithfully
 mirror real-world situations. Significantly, our results underscore the importance of two key factors: 
pedestrian flexibility and population density. These factors play a pivotal role in triggering the transition 
between mobile and immobile behavior in the steady state, a concept expounded upon within this 
paper through a straightforward order parameter known as directed mobility.
 
\end{abstract}

\end{frontmatter}

\section{Introduction}

\label{Section:Introduction}

Pedestrian dynamics \cite{Helbing-1995,Daamen,Fu2023} is a fascinating
instance of a complex system with significant implications for urban
sustainability and its planning, as highlighted in \cite{Baeza}. The
examination of patterns emerging from the counterflowing movement of
particles encompasses a broad spectrum of systems, seemingly diverse,
including pedestrian dynamics \cite%
{Tajima-2002,Ohta,Roberto-pedestrian-dynamics-2015,mestrado-eduardo-roberto2017,Chinas-que-nos-citam}
and the motion of charged colloids \cite%
{Dzubiella-2002,Rex-2007,Vissers-band-formation-2011,Buttinoni-2013}.

Surprisingly, these systems unveil more similarities than one might
initially anticipate when it comes to modeling phenomena at both micro and
macro scales. Consequently, the emergence of straight lanes and
distillation, resulting from the intricate interplay of self-propelled
and/or field-directed objects or particles, has ignited a multitude of
intriguing questions within the realms of statistical mechanics and
stochastic processes in physics. These phenomena are often simulated using
Monte Carlo (MC) simulations or expressed through Partial Differential
Equations (PDE) \cite%
{Roberto-pedestrian-dynamics-2015,mestrado-eduardo-roberto2017,nosso-PRE-clogging-2019,Stock-JSTAT-2019}%
.

Our research has delved into models considering a two-species counterflow of
particles under various circumstances, with and without particle exclusion 
\cite{Roberto-pedestrian-dynamics-2015,mestrado-eduardo-roberto2017}. More
recently, we have explored systems with exclusion, where cell occupancy
follows Fermi-Dirac statistics \cite%
{nosso-PRE-clogging-2019,Stock-JSTAT-2019}. These systems are particularly
intriguing as they reproduce compelling phenomena such as the formation of
condensates, metastability, and the coexistence of mobile and clogged states 
\cite{Stock2020}, corroborating other numerical findings \cite%
{Appert2001,Barlovic1998}.

\begin{figure}[tbph]
\begin{center}
\includegraphics[width=1.0\textwidth]{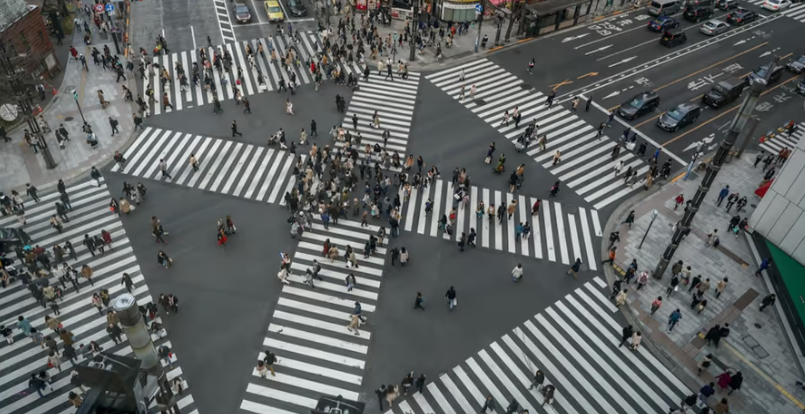}
\end{center}
\caption{A typical example of a scramble crossing in urban settings is the
bustling Ginza district in Tokyo (image source: Shutterstock). }
\label{Fig:ginza}
\end{figure}

Nevertheless, even more complex situations can occur, holding substantial
implications for the functioning of cities and urban planning. In this
context, the concept of a scramble crossing involving four distinct groups
or "species" of people striving to reach their respective objectives, as
depicted in Figure \ref{Fig:ginza}, becomes relevant. This scenario, as
observed in the Ginza district of Tokyo, provides a conspicuous example of
human congestion and its impact on urban dynamics.

In this manuscript, building upon our prior modeling work involving
two-species counterflowing particles, we present a novel model tailored to
capture scenarios exemplified by Figure \ref{Fig:ginza}. Leveraging a
parameter denoted as $\alpha $, which we previously employed in modeling
pedestrian dynamics within a nightclub environment \cite{Stock2023} to
regulate the level of objectivity or orientation of pedestrians toward their
destination, in our current context, this parameter pertains to the act of
crossing the intersection. Our objective is to investigate the
characteristic patterns that emerge at these intersections, analyze the
statistical distribution of pedestrian crossing times, and assess the impact
of pedestrian density and the degree of orientation or disorientation among
pedestrians on these statistical outcomes.

This investigation encompasses several key parameters, including the average
duration of the first crossing, its dispersion, as well as higher-order
statistics such as skewness and kurtosis for this duration. We closely
monitor snapshots of the first crossing to scrutinize the jamming patterns
of particles and their natural dispersion within this peculiar scenario.
Additionally, we track the temporal evolution of the first crossing times of
particles at various time steps.

Beyond this initial crossing analysis, we consider the presence of periodic
boundary conditions (PBC) for our particles. Following the initial crossing,
the particles continue their movement and eventually attain a steady state.
To observe and analyze this steady state after a certain number of
iterations, we employ the concept of \textit{directed mobility}---an
intriguing measure that quantifies the extent to which pedestrians move
toward their intended destinations, which, in this case, is the opposite
side of the intersection during crossings.

After the first crossing time, particles returning to their respective
corners act as new individuals entering a street, further contributing to
the dynamic behavior of the system.

The structure of our paper is outlined as follows: In the subsequent
section, we introduce our model. Section \ref{Sec:Results} is dedicated to
the exploration and analysis of our findings, and, lastly, we present our
summaries and draw conclusions in section \ref{Sec:Conclusions}.

\section{Model}

\label{Sec:Model}

\begin{figure}[tbph]
\begin{center}
\includegraphics[width=1.0\textwidth]{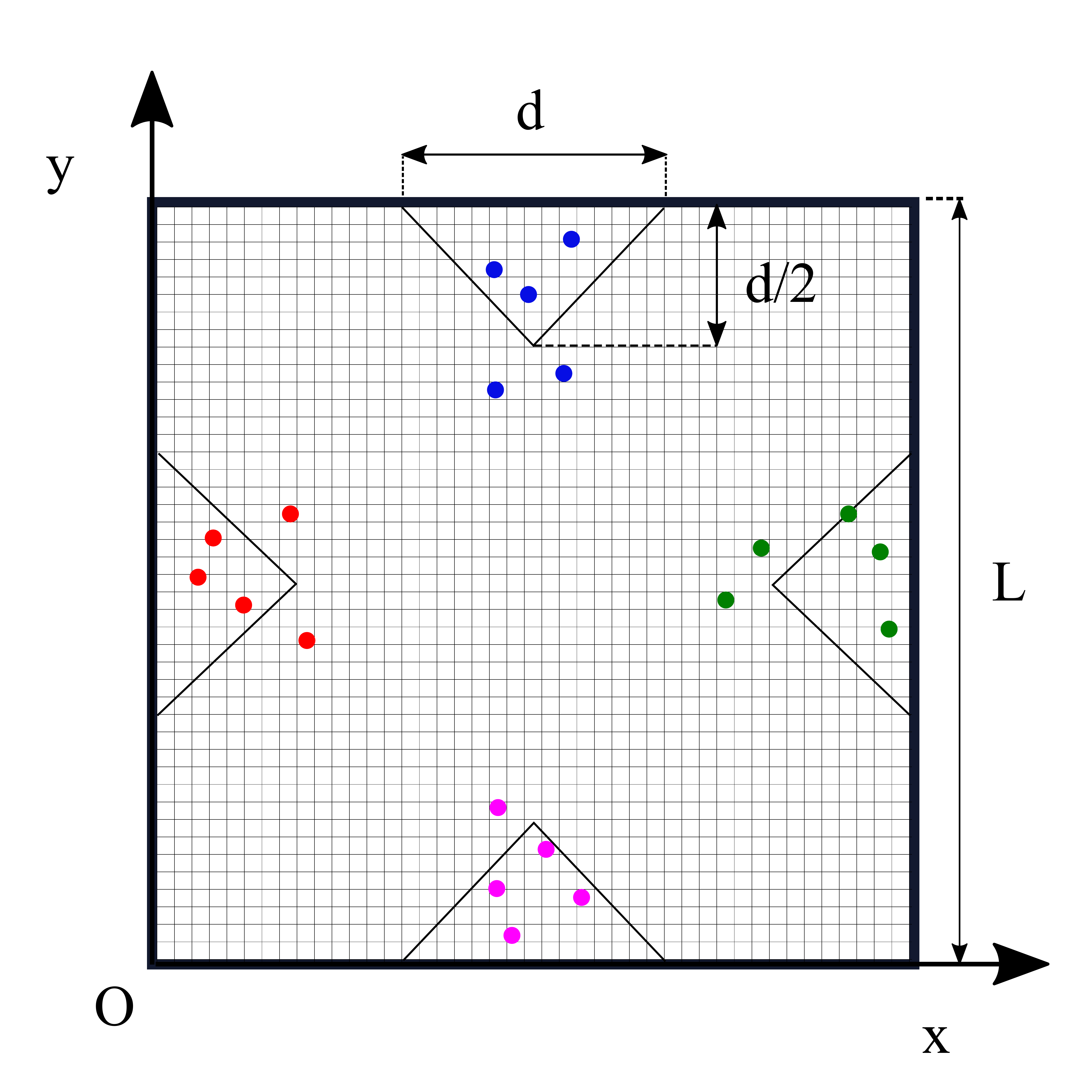}
\end{center}
\caption{Illustration of a scramble crossing featuring four corners on each
side of the lattice. Particles of each type endeavor to traverse the lattice
in pursuit of reaching the opposite sidewalk. Specifically, particles of
species $A$ are depicted in red, those of type $B$ are shown in blue,
particles belonging to type $C$ are represented in green, and particles
classified as type $D$ are denoted in pink.}
\label{depiction}
\end{figure}

We define our model as a system comprising $N$ particles, each belonging to
one of four distinct types labeled as $A$, $B$, $C$, and $D$. These
particles navigate within an underlying square lattice with dimensions $L$.
To accurately represent a scramble crossing, we delineate four sidewalk
corners by establishing four triangular regions, each centered on one side
of our square lattice, as illustrated in Figure \ref{depiction}. 
Each triangular region possesses a base measuring $d$ and a height of $d/2$.
Positioned at each sidewalk corner is a cluster of particles belonging to
the same species. These particle types are distinguished solely by their
favored direction of movement. Without sacrificing generality, we establish
that particles of type $A$ exhibit a preference for advancing in the $+x$
direction, particles of type $B$ tend to move in the $-y$ direction,
particles categorized as type $C$ gravitate towards the $-x$ direction,
while particles designated as type $D$ exhibit a tendency to proceed in the $%
+y$ direction.

The movement of particles across the lattice occurs cell by cell, governed
by a set of stochastic rules inspired by the pioneering work of Montroll and
West on clannish random walkers \cite{Montroll-stochastic-processes-1979},
subsequently adapted by da Silva et al. to simulate counterflowing
pedestrian dynamics \cite{Roberto-pedestrian-dynamics-2015}. Our approach
shares similarities with other methods, such as the static floor field model
found in the literature (see, for instance, \cite{PhysRevE.85.066128}). We
have tailored this modeling framework to accommodate our specific case
involving four distinct species. It's worth noting that our adaptation
builds upon the improved modeling techniques proposed in some of our prior
contributions \cite%
{nosso-PRE-clogging-2019,Stock-JSTAT-2019,Stock2020,Stock2023}.

The transition probabilities governing a particle's movement from cell $%
(i,j) $ to one of its neighboring cells, say, $(i+1,j)$, are as follows: 
\begin{equation}
P_{(i+1,j)\rightarrow {(i,j)}}^{(l)}=p+\alpha \left( \Delta \vec{r}_{x}\cdot 
\hat{u}\right) ,
\end{equation}%
where $\alpha $ is the coefficient regulating the degree of dynamical bias, $%
\Delta \vec{r}_{x}=\hat{e}_{x}$ represents the displacement vector, with $%
\hat{e}_{x}$ denoting the unit vector in the $x$-direction of our reference
frame, and $\hat{u}=\vec{u}/||\vec{u}||$ is the unit vector associated with
our static floor field, which guides each particle toward its destination.
Specifically, $\vec{u}=\vec{u}(\vec{r})$, where $\vec{r}=i\hat{e}_{x}+j\hat{e%
}_{y}$ denotes the particle's position in Cartesian coordinates. It's
important to note that each species within our model perceives its unique
static floor field, as their respective preferred directions of movement
differ.

As one might anticipate, the transition probabilities for the same particle
at position $(i,j)$ to move to any of its neighboring cells are defined in
relation to the displacement vector. Specifically, when the particle hops to 
$(i-1,j)$, the displacement vector is $\Delta \vec{r}_{x}=-\hat{e}_{x}$. If
it moves to $(i,j+1)$, the displacement vector is $\Delta \vec{r}_{y}=\hat{e}%
_{y}$, and if it hops to $(i,j-1)$, the displacement vector becomes $\Delta 
\vec{r}_{y}=-\hat{e}_{y}$. To ensure a normalization constraint, the
probability that a particle remains in its current cell is expressed as: 
\begin{equation}
P_{(i,j)\rightarrow {(i,j)}}^{(l)}=1-\sum\limits_{\left\langle i\prime
,j\prime \right\rangle }Pr_{(i,j)\rightarrow {(i\prime ,j\prime )}%
}^{(l)}=1-4p\text{.}
\end{equation}

To assess the degree of dynamism inherent in the particle's pursuit of
reaching the opposite corner, we define \textit{directed mobility} as 
\begin{equation}
M_{\vec{u}}^{(l)}\equiv \frac{m_{\vec{u}}^{(l)}}{N},  \label{Eq:Mobility}
\end{equation}%
where $m_{\vec{u}}^{(l)}$ represents the total number of particles of any
kind (all four species are considered) that have moved to a neighboring cell
to which their species' static floor field had a vectorial component
pointing towards.

We implemented our model through Monte Carlo (MC) simulations, employing an
asynchronous update scheme for particle positions. Under this asynchronous
updating scheme, during each MC time step, we randomly select $N$ particles
from the total $N$ particles constituting our system and update their
positions sequentially based on the transition probabilities defined earlier.

We will systematically organize our results into two subsections within the
upcoming section. Firstly, we strategically placed four groups of particles
at the four different corners with carefully chosen initial conditions. Our
initial focus lies in analyzing the time required for these particles to
make their first crossing of the system.

Secondly, given our implementation of periodic boundary conditions (PBC) for
our particles, following this initial crossing, the particles continue to
move and ultimately attain a steady state. In this stable configuration,
particles return to their respective corners, reminiscent of new individuals
entering a street. We delve into this steady state by monitoring the average
mobility of the system.

The parameters characterizing the steady-state mobility reveal a transition
between a situation where the system remains mobile and one where it becomes
congested. This transition is evident in both parameters: $\alpha $, which
governs the level of objective orientation of pedestrians toward their
destination, and $\rho $, representing the density of pedestrians, with our
results suggesting a relationship between these two parameters in such
\textquotedblleft crossover\textquotedblright\ situation.

\section{Results}

\label{Sec:Results}

We beging our presentation of results by considering an initial condition
that mimics a "real-world" scenario. In this setup, all particles are
initially positioned at each corner of the four-way scramble crossing, or
within the triangular region situated in the middle of each side of our
square lattice (as depicted in Fig. \ref{depiction}). In technical terms,
this entails that particles of types $A$, $B$, $C$, and $D$ are initially
distributed across sites selected at random, subject to the following
constraints: 
\begin{equation*}
x+\frac{L-d}{2}\leq y\leq -x+\frac{L+d}{2},\quad \mbox{for}\quad 0\leq x\leq
d/2,
\end{equation*}%
\begin{equation*}
-y+\frac{3L-d}{2}\leq x\leq y-\frac{L-d}{2},\quad \mbox{for}\quad L-d/2\leq
y\leq L,
\end{equation*}%
\begin{equation*}
-x+\frac{3L-d}{2}\leq y\leq x-\frac{L-d}{2},\quad \mbox{for}\quad L-d/2\leq
x\leq L,
\end{equation*}%
and 
\begin{equation*}
y-\frac{L-d}{2}\leq x\leq -y+\frac{L+d}{2},\quad \mbox{for}\quad 0\leq y\leq
d/2,
\end{equation*}%
respectively.

Under the given initial conditions, we embarked on a study of the time
required for each agent to complete a single crossing---specifically,
reaching either the opposing sidewalk corner or breaching the lattice
boundaries (defined as beyond the limits of $1\leq x\leq L$ and $1\leq y\leq
L$).

In this preliminary analysis, our primary emphasis is directed towards the
concept of \textquotedblleft first crossing,\textquotedblright\ a topic we
explore in greater detail in the subsequent subsection. Following this
initial investigation, we shift our focus towards scrutinizing the
parameters during the steady state, which naturally ensues after several
iterations. As mentioned earlier, this steady state simulates a scenario
where pedestrians have the green light to cross, and those entering the
street assume the roles of new pedestrians.

\subsection{First time crossing}

\label{one_cross}

\begin{figure}[t]
\centering%
\begin{subfigure}[t]{0.475\textwidth}
\centering
\includegraphics[width=\textwidth]{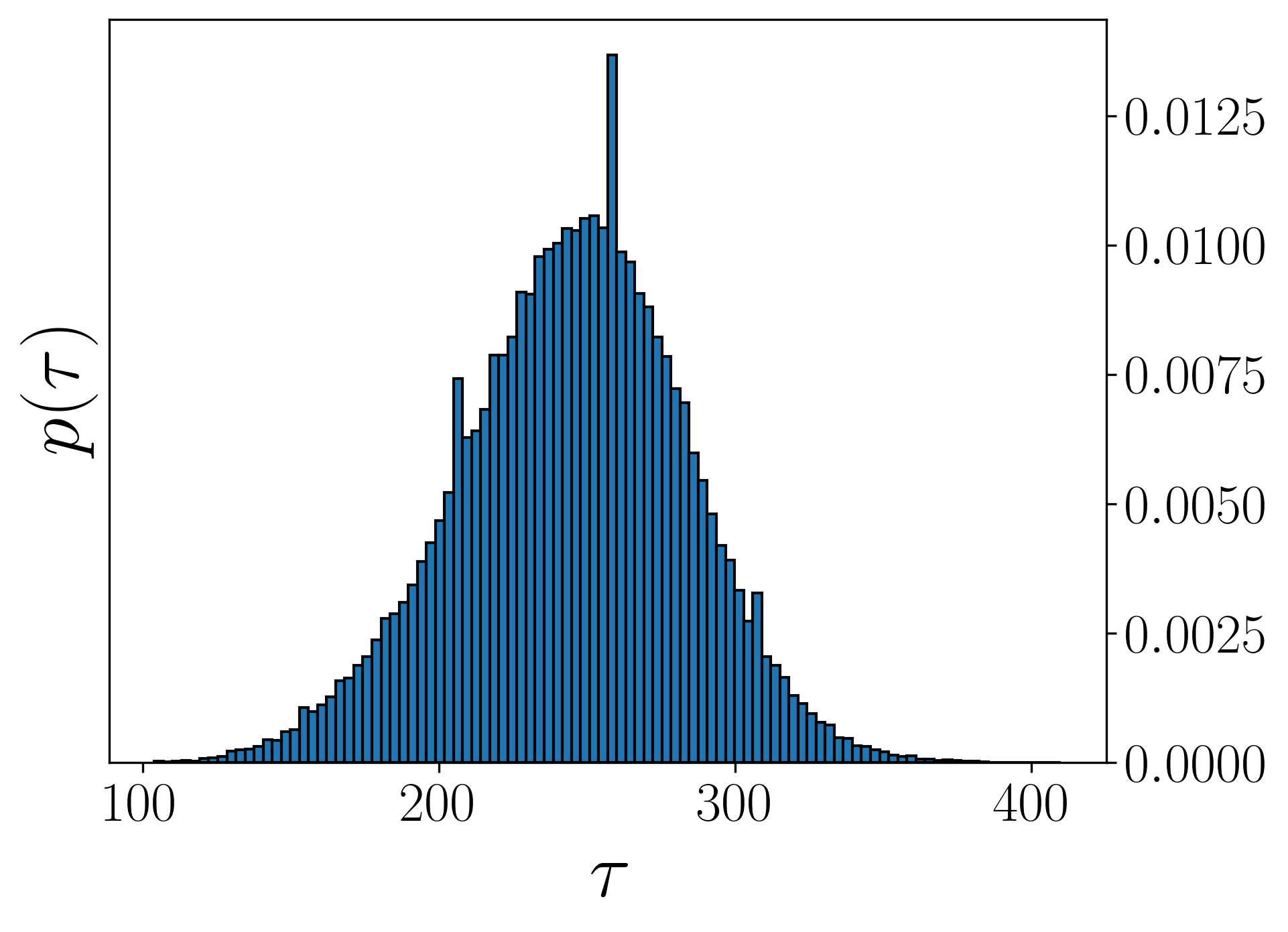} 
\caption{$\rho=0.03$.} \label{fig:snap_1}
\end{subfigure}\hfill 
\begin{subfigure}[t]{0.475\textwidth}
\centering
\includegraphics[width=\textwidth]{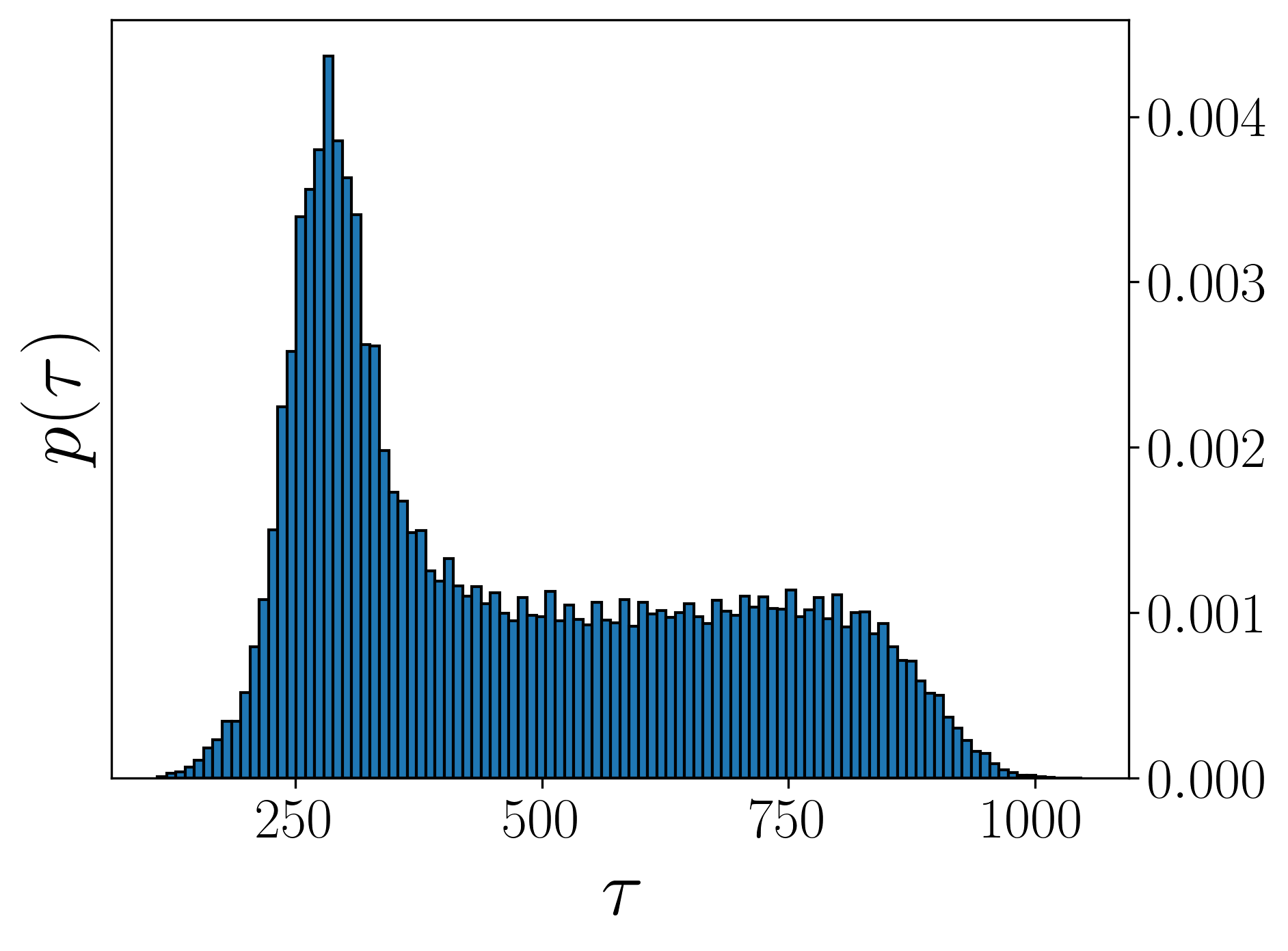} 
\caption{$\rho=0.06$.} \label{fig:snap_2}
\end{subfigure}\vskip\baselineskip%
\begin{subfigure}[t]{0.475\textwidth}
\centering
\includegraphics[width=\textwidth]{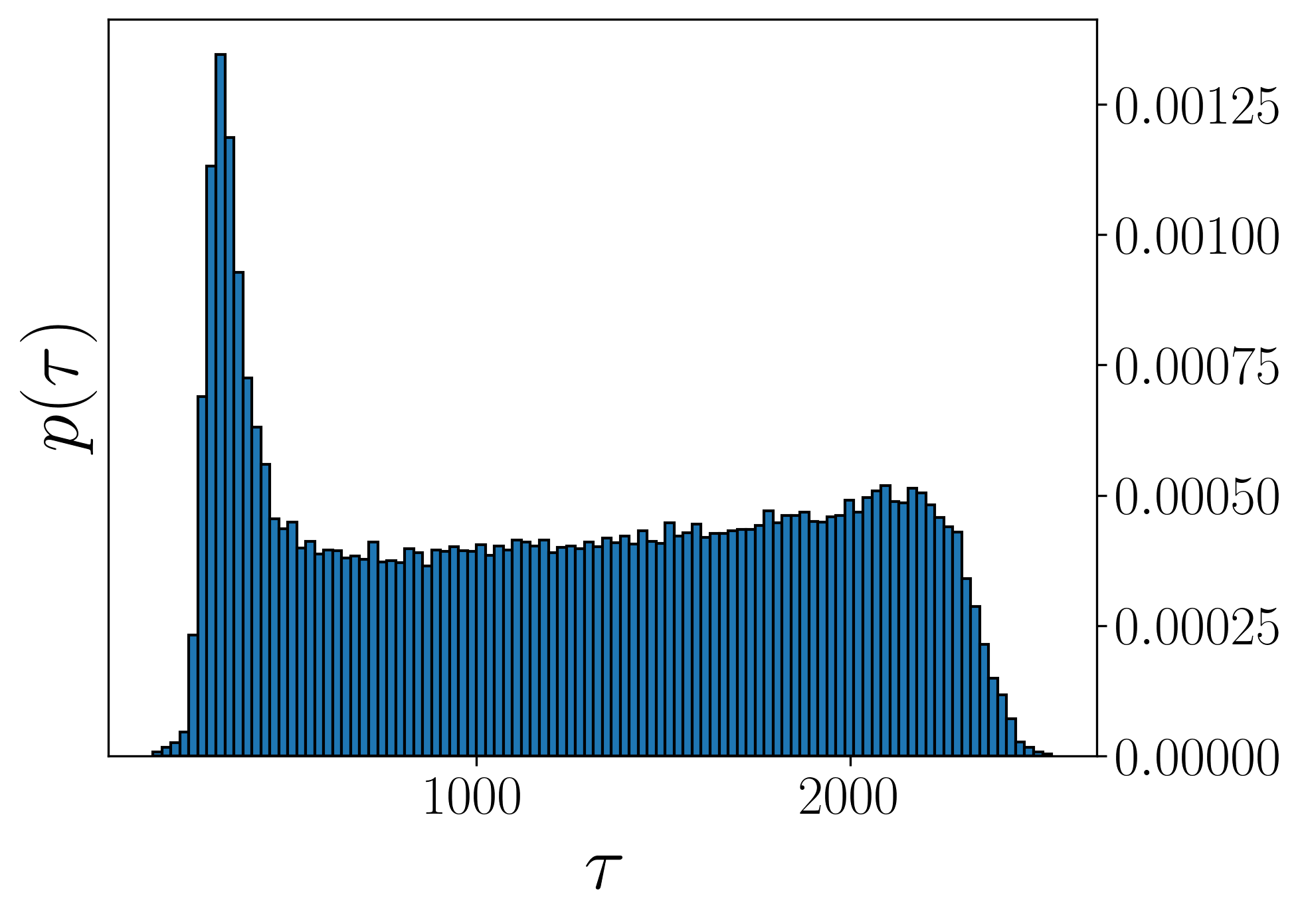} 
\caption{$\rho=0.12$.} \label{fig:snap_3}
\end{subfigure}\hfill 
\begin{subfigure}[t]{0.475\textwidth}
\centering
\includegraphics[width=\textwidth]{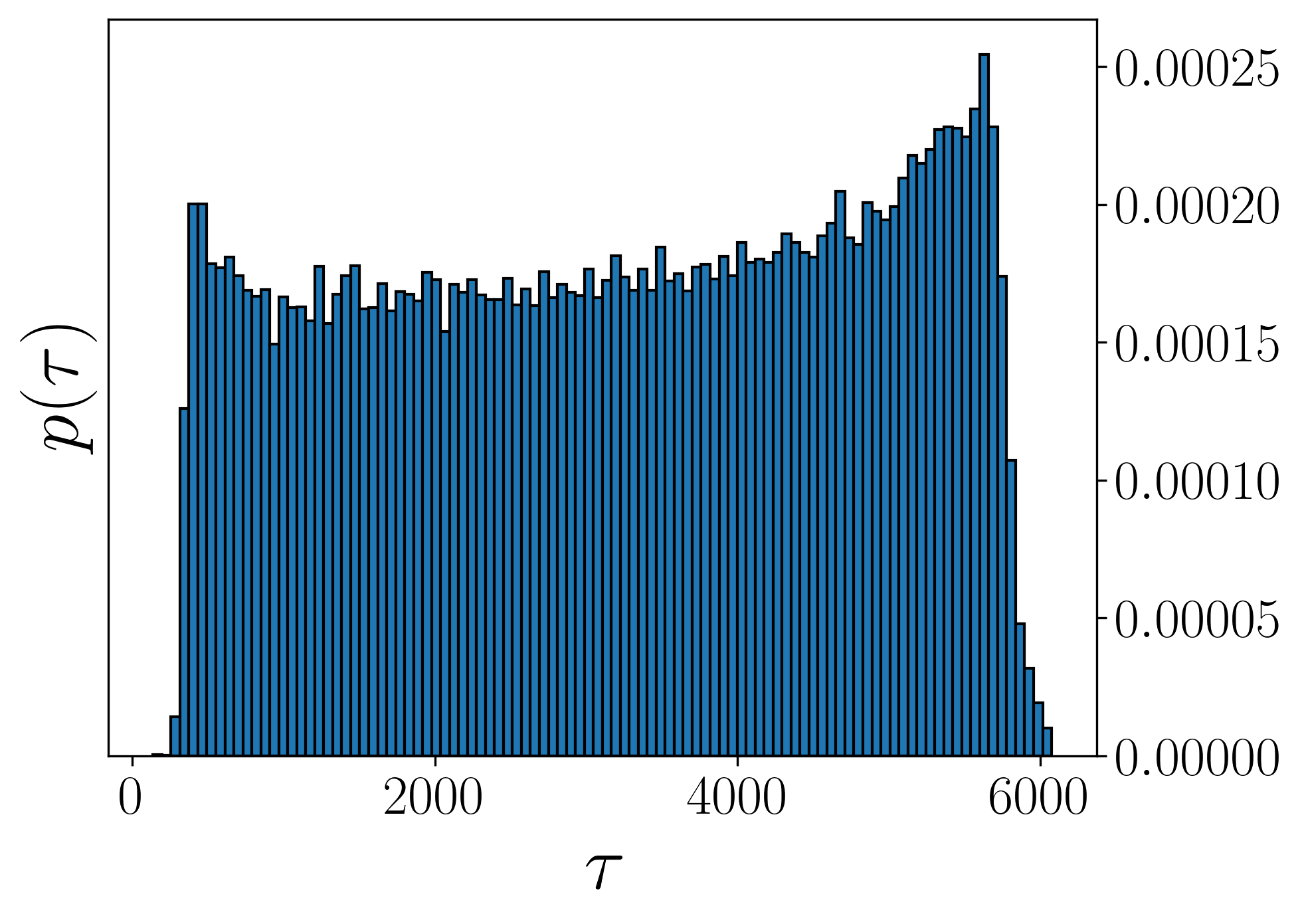} 
\caption{$\rho=0.24$.} \label{fig:snap_3}
\end{subfigure}
\caption{The distribution of first crossing times for various density values
within a system characterized by $L=128$, $d=64$, and $\protect\alpha =0.249$%
.}
\label{distro_cbc}
\end{figure}

In Figure \ref{distro_cbc}, we present the first crossing time distribution,
taking into account the initial conditions previously specified, across various
particle density values. Our simulations encompass a system of dimensions $%
L=128$, with a corner dimension of $d=64$, and a parameter $\alpha $ set to $%
0.249$.

Upon examination, it becomes evident that in scenarios characterized by low
particle density, the distribution exhibits characteristics resembling a
normal distribution, albeit with a slightly negative kurtosis. However, as
we transition to higher density scenarios, the distribution undergoes a
notable transformation. It begins to exhibit an increase in its root mean
square error until reaching a critical point where we observe a more uniform
distribution pattern with a broader support. In fact, this transition to
higher density scenarios can lead to both shorter and longer first crossing
times, a phenomenon attributed to congestion resulting from increased
density levels. 

The phenomenon of "uniformization" within the distribution of first crossing
times becomes clearer when we examine snapshots of the spatial distribution
of particles at various time points during the simulation. In Figure \ref%
{snapshots_cbc}, we can discern the following trends:

\begin{enumerate}
\item Initially, at$\ t=17$ MC steps (depicted in plot (a)), only a limited
number of particles from each group disengage from the initial cluster.

\item Over time, around $t=622$ MC steps (illustrated in plot (b)), a
substantial cluster of particles congregates in the central region of the
scramble crossing. This outcome is unsurprising given the high particle
density within the system.

\item Furthermore, we observe that a smaller contingent of particles from
each species manages to encircle the primary cluster, successfully advancing
towards their respective objectives. This seemingly consistent rate at which
particles encircle the central cluster plays a crucial role in achieving a
uniform distribution of crossing times.

\item Ultimately, after a sufficient period has passed, the main cluster
disintegrates, leaving only small groups of particles on their individual
journeys toward their destinations.
\end{enumerate}

This series of observations sheds light on the dynamic processes at play and
how they contribute to the even distribution of crossing times. 
\begin{figure}[t]
\centering
\begin{subfigure}[t]{0.31\textwidth}
\centering
\includegraphics[width=\textwidth,trim={12cm 12cm 12cm 12cm},clip]{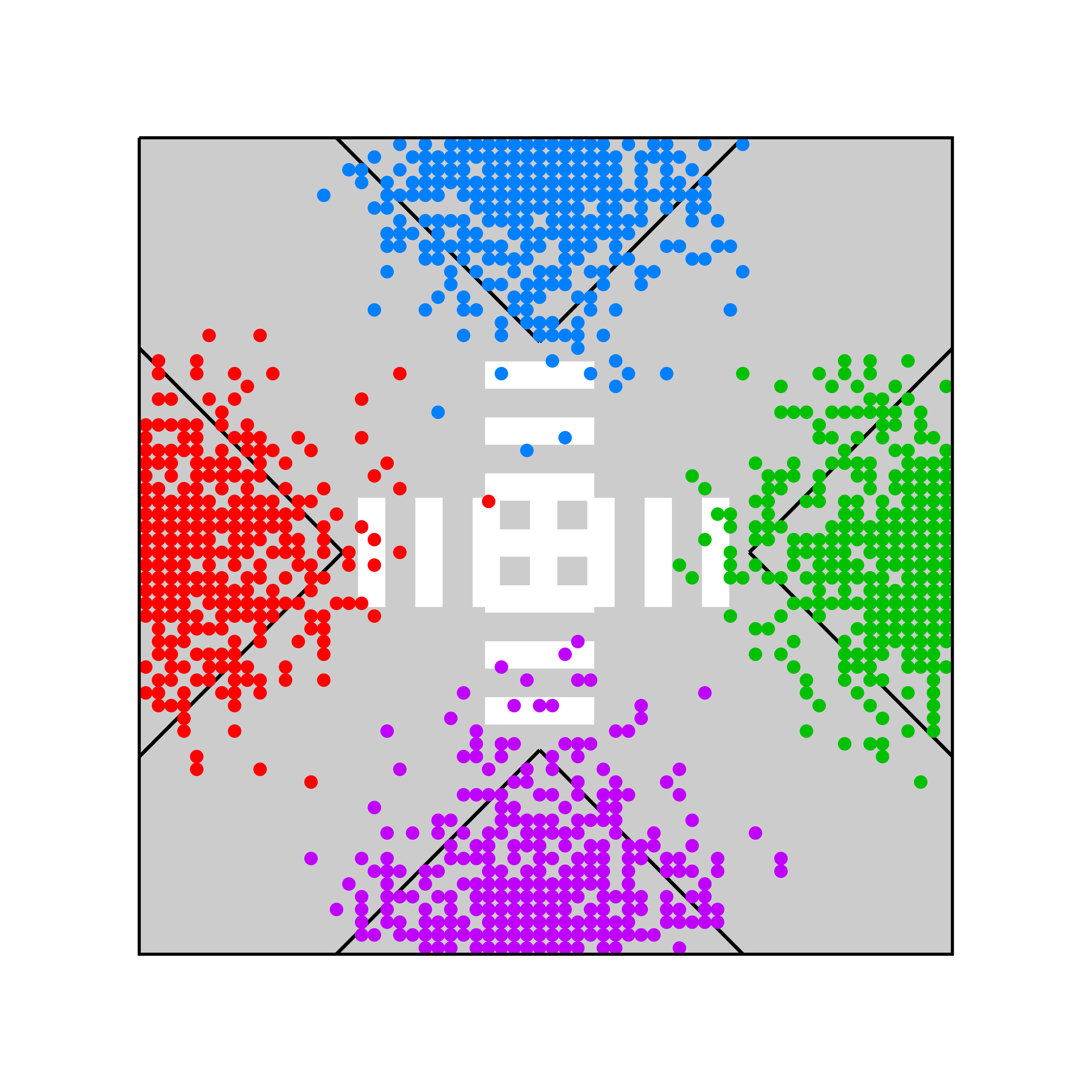} 
\caption{Snapshot for $t=17$ MC steps.} \label{fig:snap_1}
\end{subfigure}
\hfill 
\begin{subfigure}[t]{0.31\textwidth}
\centering
\includegraphics[width=\textwidth,trim={12cm 12cm 12cm 12cm},clip]{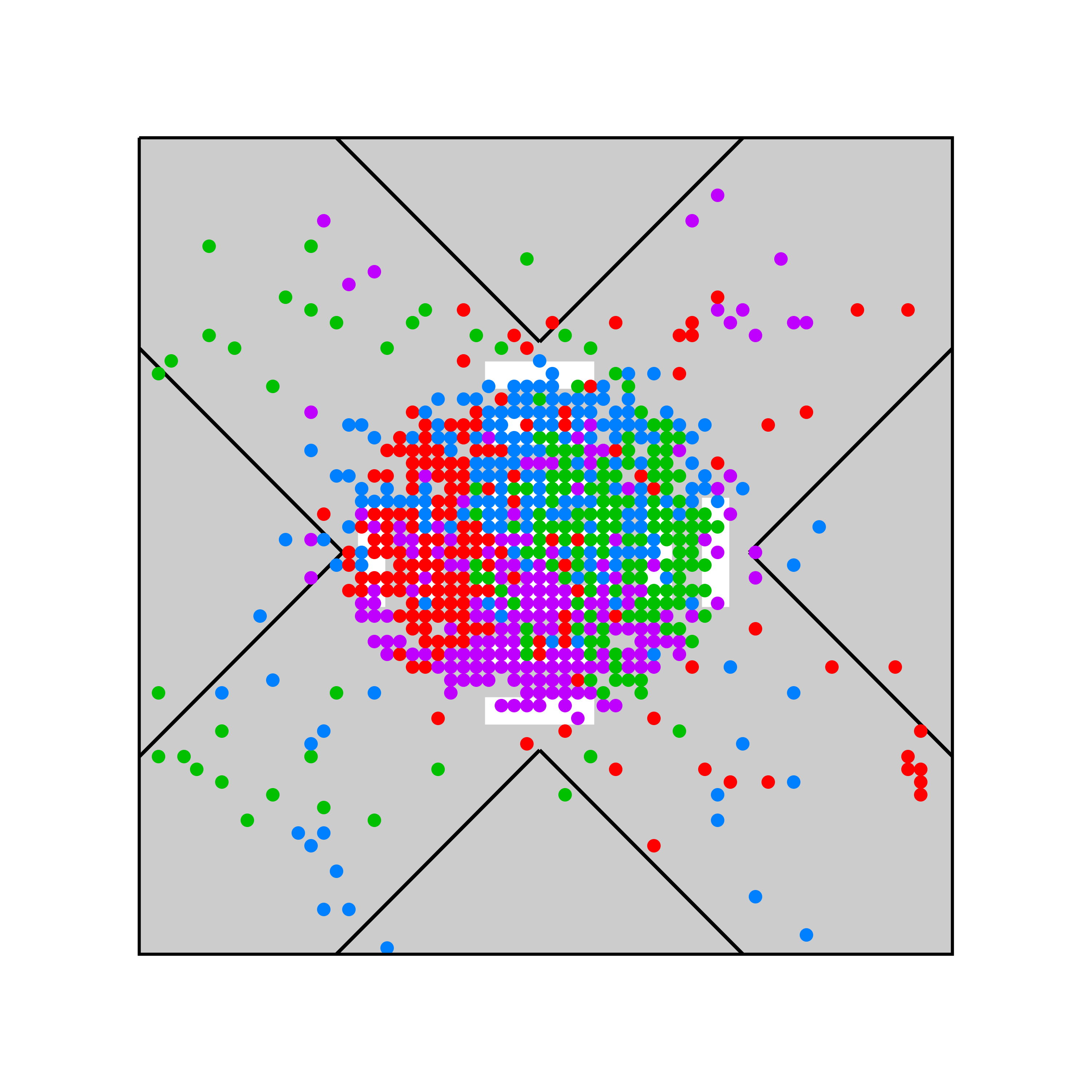} 
\caption{Snapshot for $t=622$ MC steps.} \label{fig:snap_2}
\end{subfigure}
\hfill 
\begin{subfigure}[t]{0.31\textwidth}
\centering
\includegraphics[width=\textwidth,trim={12cm 12cm 12cm 12cm},clip]{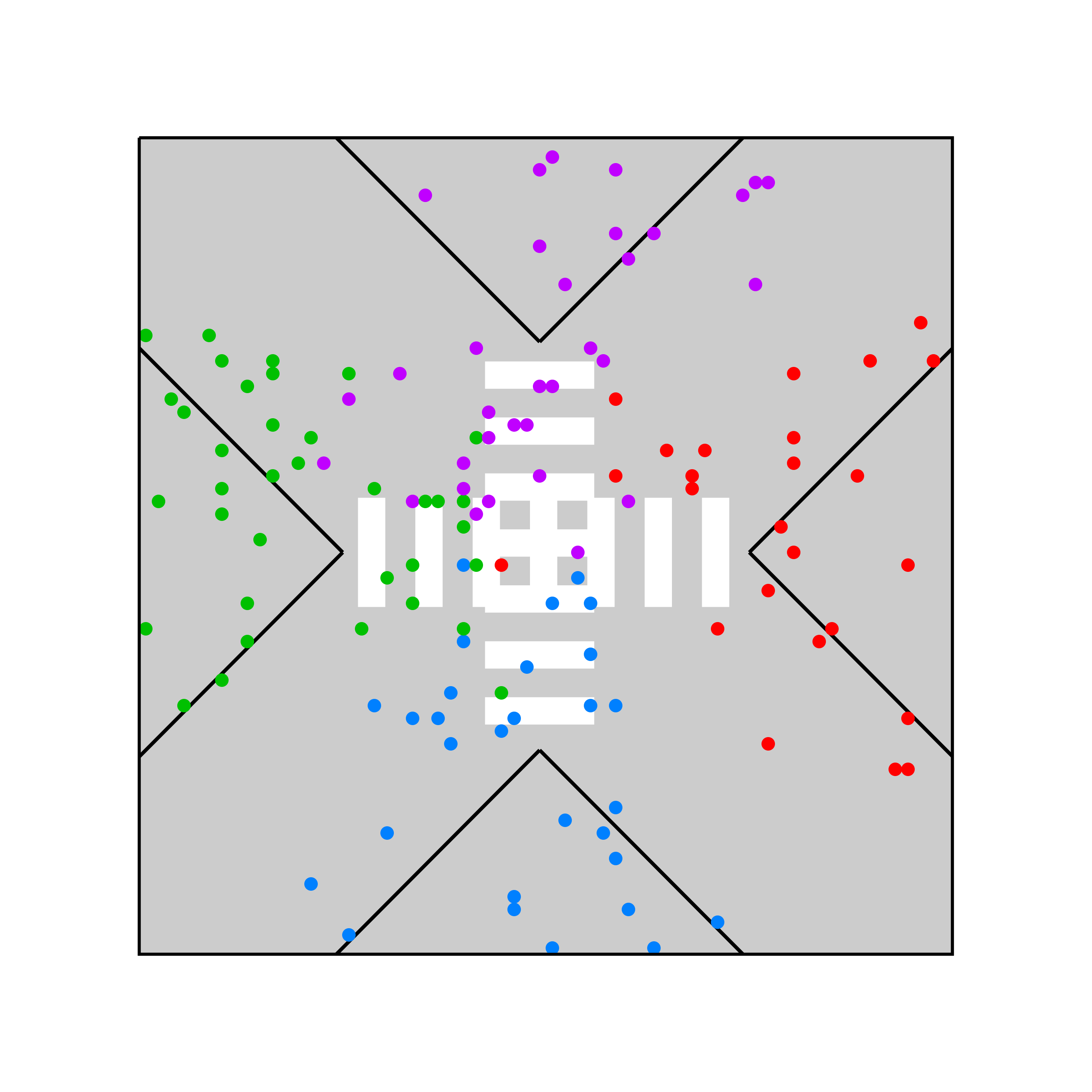} 
\caption{Snapshot for $t=1187$ MC steps.} \label{fig:snap_3}
\end{subfigure}
\caption{Sequential snapshots depicting the particle distribution at
distinct time intervals in a system characterized by $L=128$, $d=64$, $%
\protect\rho =0.25$, and $\protect\alpha =0.249$. In the initial snapshot
(plot (a)), select particles detach from the primary group, which remains
clustered at the corner. As time advances (plot (b)), a substantial central
cluster emerges at the crossing, coexisting with particles employing
flanking strategies to reach their destinations. Ultimately, in the final
snapshot, the central cluster disperses while some particles continue their
journey across the intersection.}
\label{snapshots_cbc}
\end{figure}

Returning to Fig. \ref{distro_cbc}, we can discern distinct patterns in the
crossing time distribution, particularly when examining the impact of
varying particle densities. In Fig. \ref{distro_cbc} (a), where the particle
density is lower ($\rho =0.03$), we observe a bell-shaped distribution
reminiscent of a normal curve, with an approximate mean value of $%
\left\langle \tau \right\rangle \approx 250$ MC steps and a standard
deviation of $\sigma_{\tau } \approx 50$ MC steps.

Moving on to intermediate density values within the range of $0.03<\rho
<0.24 $, as depicted in Figs. \ref{distro_cbc} (b) and \ref{distro_cbc} (c), we notice a notable
shift. While the main peak of the distribution remains consistent, there is
now an elongated right tail extending from the peak, resulting in the
emergence of a plateau-like feature. This phenomenon aligns with the
presence of the central cluster, as observed in Fig. \ref{snapshots_cbc} (b).
The central cluster fosters a steady influx of particles toward the opposing
corner, contributing to the uniform distribution pattern we observe in the
crossing times.

\begin{figure}[tbph]
\begin{center}
\includegraphics[width=1.0\textwidth]{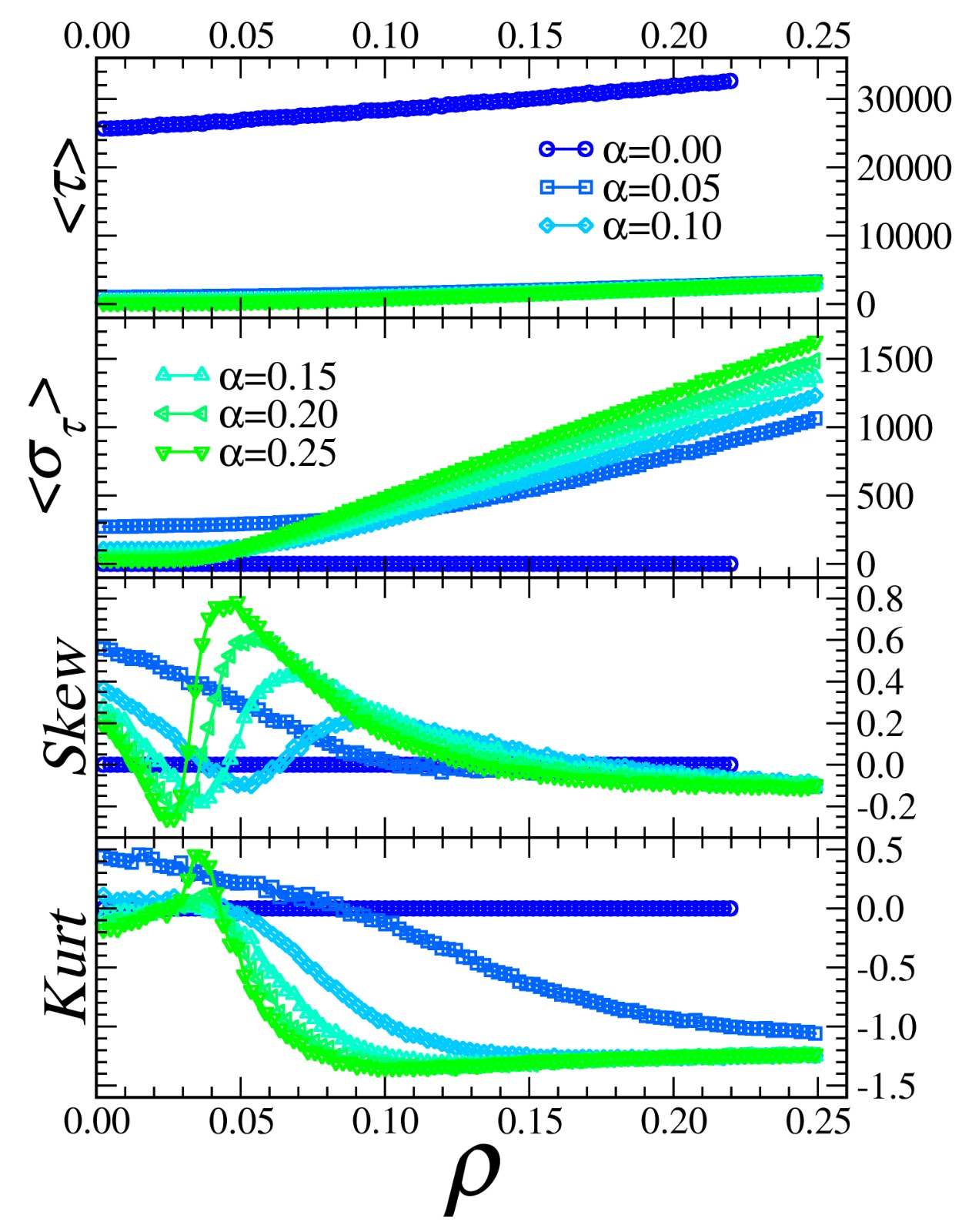}
\end{center}
\caption{Analyzing crossing time statistics across various $\protect\alpha $
values as a function of the density of pedestrian $\protect\rho $.}
\label{Fig:statistics_cbc}
\end{figure}

Given this behavior, it becomes intriguing to investigate the fluctuations
in the first crossing time as a function of $\rho $ across various $\alpha $
values, as illustrated in Fig. \ref{Fig:statistics_cbc}. To do so, we performed $N_{run}$ simulations (each with a different seed) to obtain a sample size of $m=10^5$ crossing times, such that the relation $N*N_{run}=m$ was maintained.

The average first crossing time is notably influenced by the parameter $%
\alpha $. A minor increase in this orientation parameter leads to a decrease
in the average time, with a lesser dependence on $\rho $, as evident from
our observations. In contrast, the dispersion, denoted as $\sigma _{\tau }$
and calculated as the difference between the second moment $\left\langle
\tau ^{2}\right\rangle $ and the square of the first moment $\left\langle
\tau \right\rangle ^{2}$, exhibits a strong dependence on particle density
for $\alpha >0$. This finding aligns with the observations presented in Fig. %
\ref{distro_cbc}. Furthermore, other significant statistical measures,
including higher moments of the first-time crossing distribution, such as
skewness:

\begin{equation}
skew=\left\langle \left( \frac{\tau -\left\langle \tau \right\rangle }{%
\sigma _{\tau }}\right) ^{3}\right\rangle \text{,}
\end{equation}%
and the heaviness of the distribution tail which is quantified by the excess
kurtosis, obtained by subtracting 3 from the fourth central moment 
\begin{equation}
kurt=\left\langle \left( \frac{\tau -\left\langle \tau \right\rangle }{%
\sigma _{\tau }}\right) ^{4}\right\rangle -3,
\end{equation}%
also present a substantial dependency on density for $\alpha >0$. The key
takeaway is that, regardless of the pedestrians' orientation levels, if it
is only positive, we consistently observe a significant influence on the
shape of the first crossing distribution.

Now, we will investigate the dynamics under periodic boundary conditions to
monitor mobility (as defined in Eq. \ref{Eq:Mobility}) in the steady state
as a function of $\alpha $ and $\rho $.

\subsection{Periodic boundary conditions}

\label{periodic}

\begin{figure}[h!]
\centering%
\begin{subfigure}[t]{0.475\textwidth}
\centering
\includegraphics[width=\textwidth]{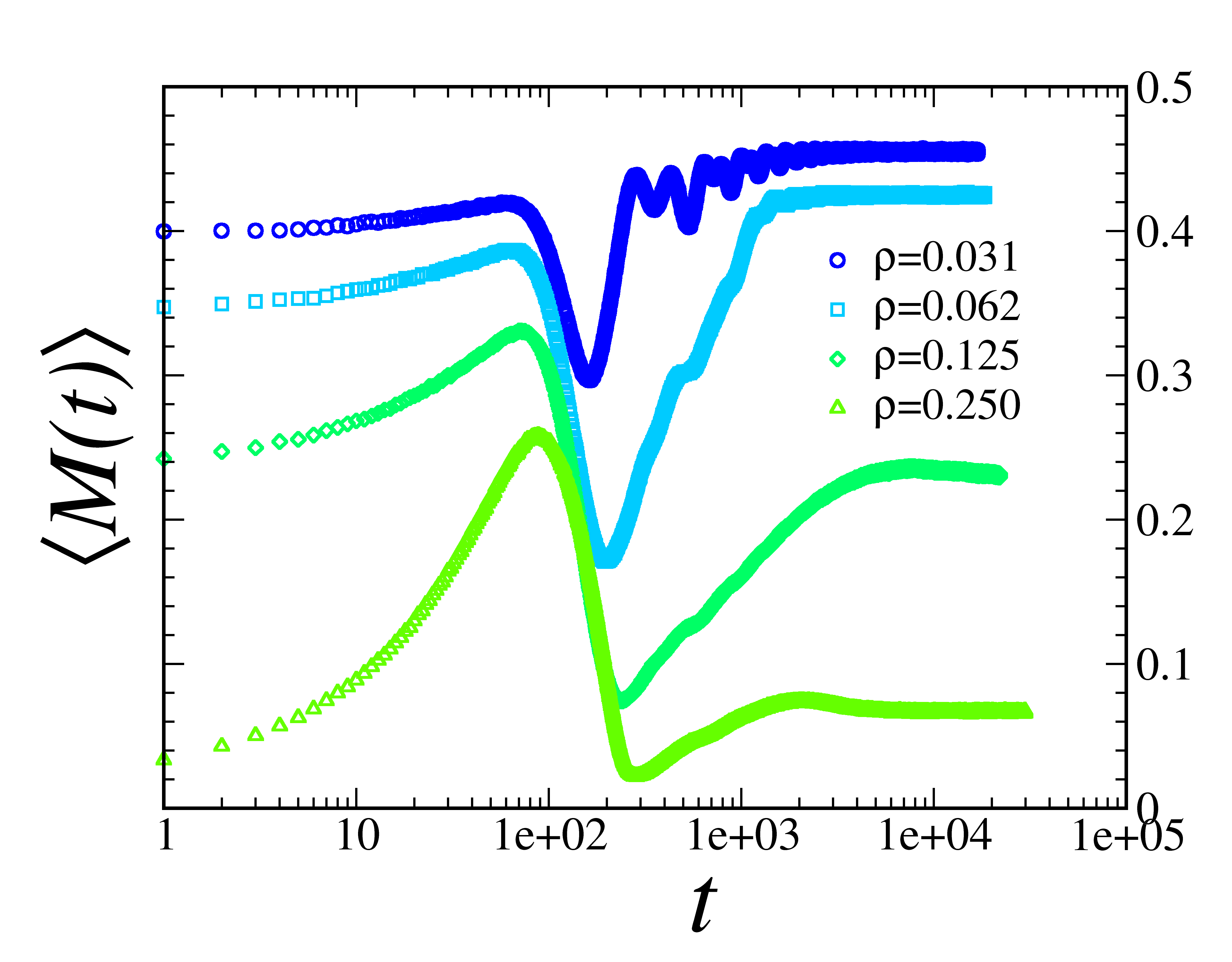} 
\caption{Time series of the average mobility for different densities, with a fixed value of $\alpha=0.2$}
 \label{fig:snap_2}
\end{subfigure}\hfill 
\begin{subfigure}[t]{0.475\textwidth}
\centering
\includegraphics[width=\textwidth]{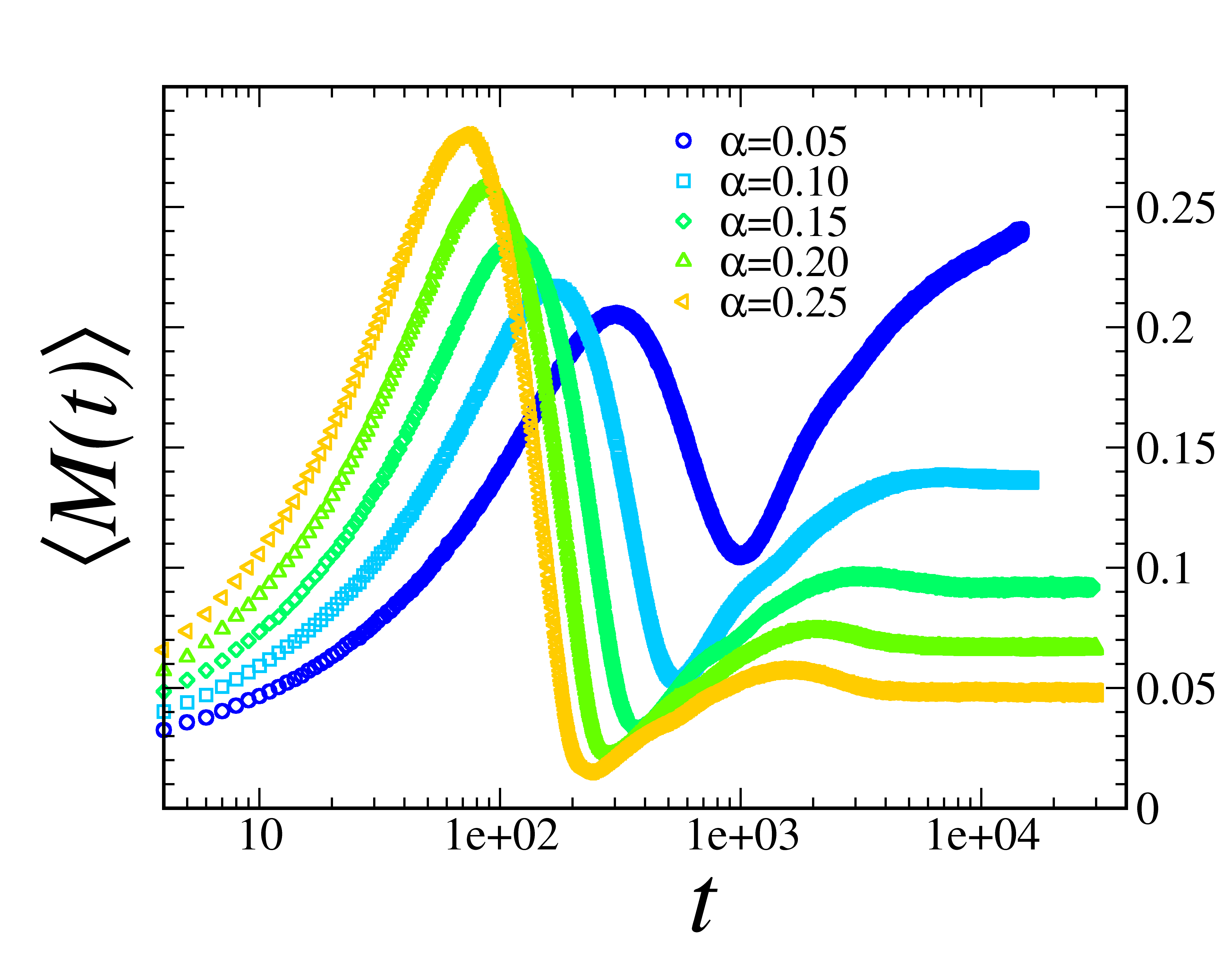} 
\caption{Time series of the average mobility for different values of $\alpha$, while keeping $\rho$ fixed at 0.25.}
\label{fig:snap_3}
\end{subfigure}
\caption{Time series illustrating the averaged directed mobility across $N_{run}=100$ samples.}
\label{Fig:Mobility_PBC}
\end{figure}

We now turn our attention to analyzing the dynamics of intersection
crossing, taking into account periodic boundary conditions (PBC). Initially,
we investigate the time-dependent behavior of average mobility ($%
\left\langle M\right\rangle $ obtained from $N_{run}=100$ different time
series) across various densities, as depicted in Fig. \ref{Fig:Mobility_PBC}
(a). In this analysis, we maintain $\alpha $ at a fixed value of $0.2$.
Notably, we observe mobility undergoing fluctuations, both increasing and
decreasing, before eventually reaching a steady state, the characteristics
of which are density-dependent.

The substantial reduction in mobility corresponds to the initial competition
among particles, as they contend with counterflow and congestion. As
depicted in Fig. \ref{Fig:Mobility_PBC} (b), the average mobility varies
across different values of $\alpha $. Notably, it is crucial to observe that
for lower orientation levels (lower $\alpha $ values), the attainment of a
steady state is anticipated to require a significantly longer time, in line
with our intuitive expectations.

\begin{figure}[t]
\centering%
\begin{subfigure}[t]{0.475\textwidth}
\centering
\includegraphics[width=\textwidth]{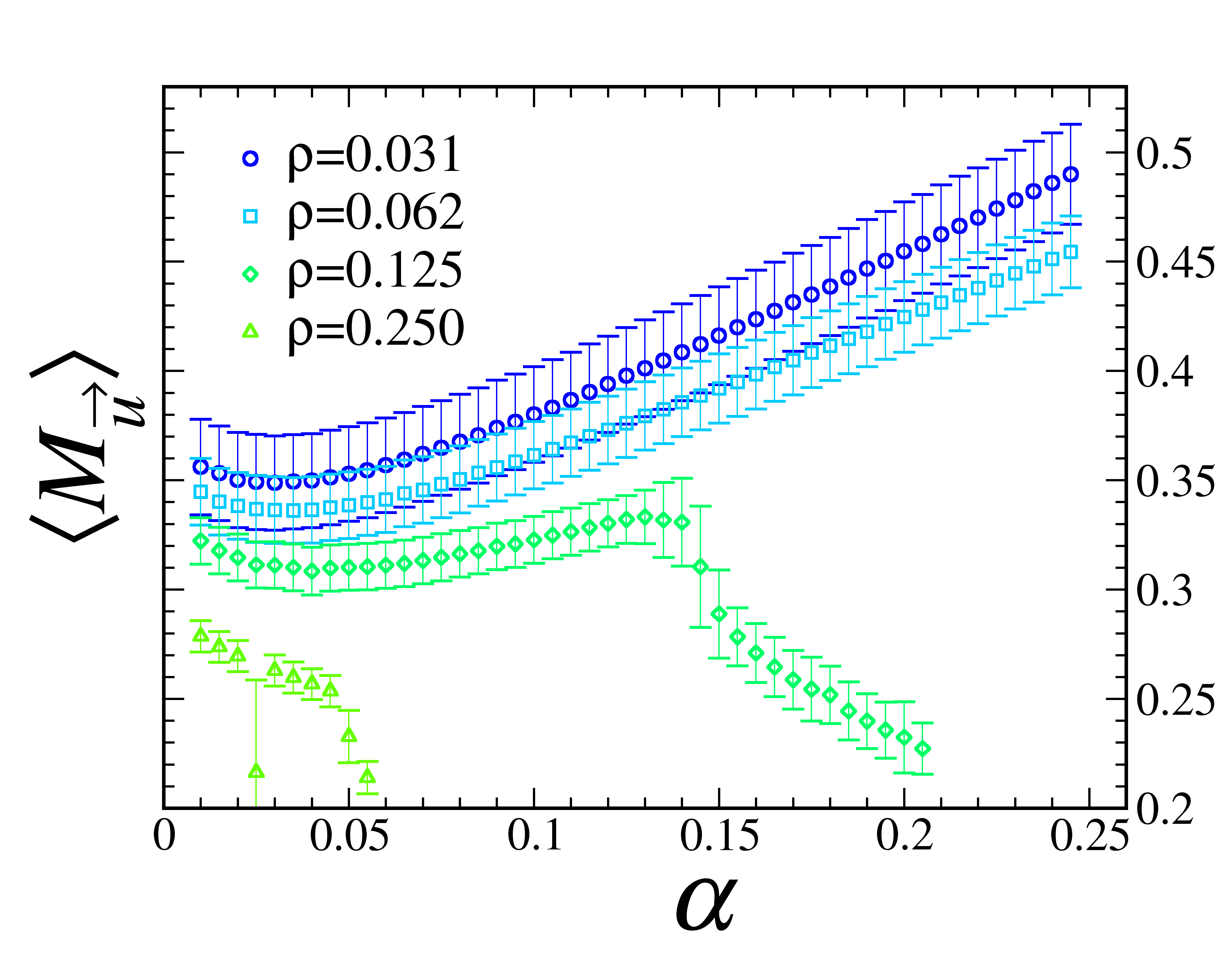} 
\caption{Steady-state average mobility as a function of $\alpha$ across various density values. }
 \label{fig:snap_2}
\end{subfigure}\hfill 
\begin{subfigure}[t]{0.475\textwidth}
\centering
\includegraphics[width=\textwidth]{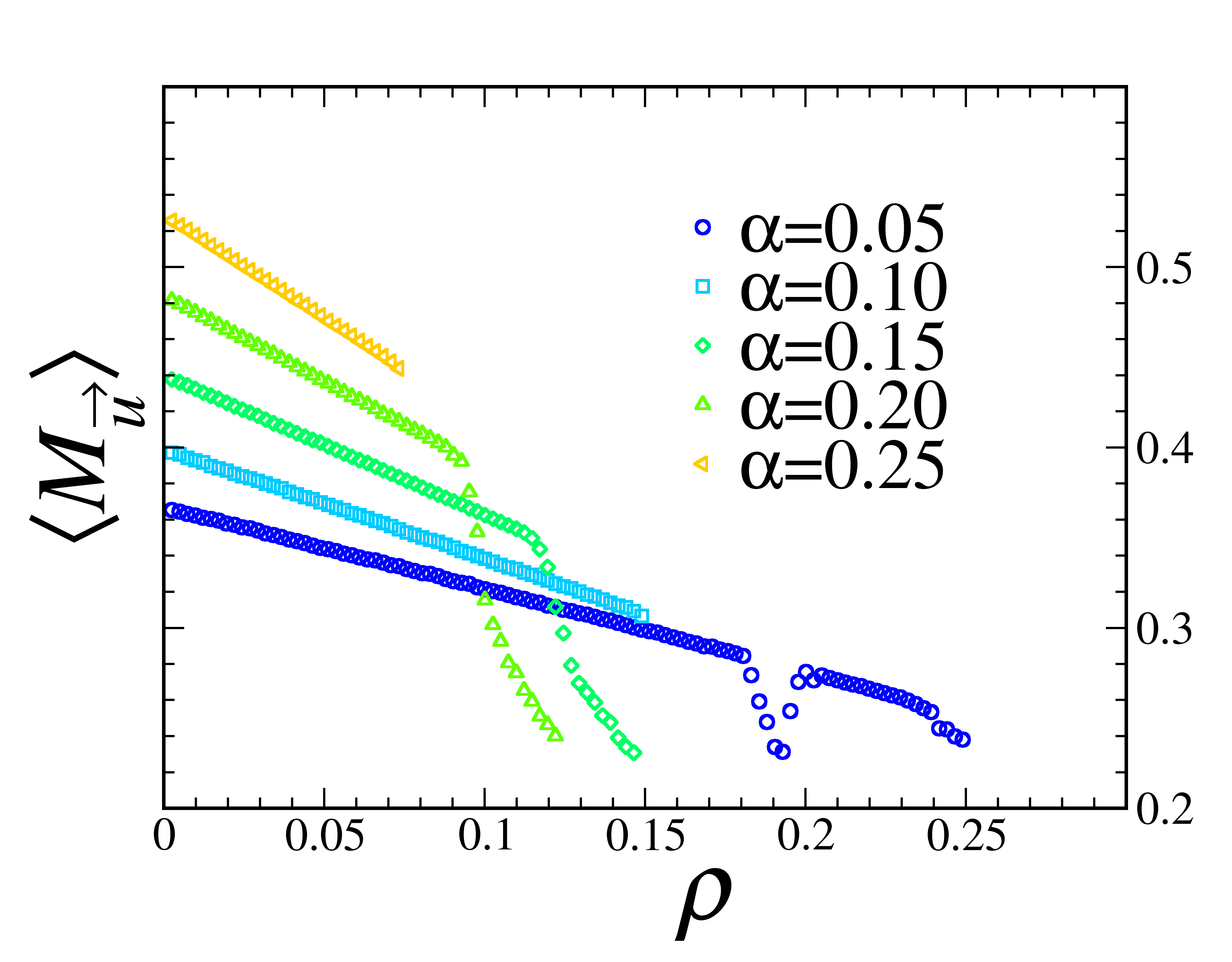} 
\caption{Steady-state average mobility as a function of $\rho$ across different values of $\alpha$}
\label{fig:snap_3}
\end{subfigure}
\caption{We note that higher values of $\protect\alpha $ enhance mobility
for lower densities ($\protect\rho <0.125$). However, beyond a certain
threshold, specifically within the range of $0.062<\protect\rho <0.125$, we
observe a decrease in mobility when particles exhibit strong oriented
movement (no flexible pedestrians)}
\label{Fig: mobility_pbc_steady_state}
\end{figure}

We now delve into an analysis of the average directed mobility in the
steady state, examining their dependency on the parameters $\alpha $ and $%
\rho $. Specifically, we explore the impact of varying $\rho $ values in the
first part and different $\alpha $ values in the second. These findings are
presented in Figs. \ref{Fig: mobility_pbc_steady_state} (a) and \ref{Fig: mobility_pbc_steady_state} (b),
respectively, where we performed $N_{run}$ simulations observing the relation $N*N_{run}=m$ for a sample size of $m=10^5$.

In Fig. \ref{Fig: mobility_pbc_steady_state} (a), we observe that by
adjusting the density ($\rho $), we can anticipate a transition between a
mobile and an immobile phase of particles, characterized by a critical value 
$\alpha _{C}(\rho)$. This transition highlights that high orientation, which
represents pedestrians with lower flexibility, tends to lead to congestion
scenarios.

Similarly, in \ref{Fig: mobility_pbc_steady_state} (b), we identify a
critical density $\rho _{C}(\alpha )$ that determines the transition between
mobile and immobile phases as $\alpha $ varies.

\section{Summaries, conclusions, and discussions}

\label{Sec:Conclusions} 
\qquad We have gathered valuable insights into pedestrian crossing times
within the intricate settings of typical urban corners in large cities. Our
results reveal instances of congestion that arise not only during the
initial competitive phase among pedestrians but also persist in the steady
state. This congestion is contingent on the interplay between two crucial
factors: the orientation level $\alpha $ and the pedestrian density $\rho $.
Our findings underscore that the flexibility and concentration of
pedestrians play pivotal roles in facilitating smooth crossings at complex
intersections in bustling urban centers.

Furthermore, we have observed a noteworthy transition between mobile and
immobile phases during the steady state, a transition that hinges on the
values of both $\alpha $ and $\rho $. It is worth emphasizing that the
concept of direct mobility has proven to be an invaluable parameter for
analyzing both transient and steady-state dynamics in pedestrian behavior
within such complex environments.

\section*{CRediT authorship contribution statement}

All authors conceived and designed the analysis, performed formal analysis,
wrote the paper, elaborated the algorithms, analysed the results, and
reviewed the manuscript.

\section*{Declaration of competing interest}

The authors declare that they have no known competing financial interests or
personal relationships that could have appeared to influence the work
reported in this paper

\section*{Funding}

R. da Silva would like to thank CNPq for financial support under grant
number 304575/2022-4. E. V. Stock also thanks CNPq for financial support
under grant numbers 152715/2022-3.

\end{document}